\documentclass[a4paper,UKenglish,cleveref, autoref, thm-restate]{lipics-v2021}
\pdfoutput=1 %uncomment to ensure pdflatex processing (mandatatory e.g. to submit to arXiv)
\hideLIPIcs  %uncomment to remove references to LIPIcs series (logo, DOI, ...), e.g. when preparing a pre-final version to be uploaded to arXiv or another public repository

\newcommand{\Jq}[1]{\textcolor{red}{\textbf{Junqi:} #1}}
\usepackage{microtype}
\usepackage{upgreek}
\usepackage{amsmath, amssymb}
\usepackage{tcolorbox}

\title{Faster exponential algorithms for cut problems via geometric data structures} %TODO Please add

\author{L\'{a}szl\'{o} Kozma}{Institut für Informatik, Freie Universität Berlin, Germany}{laszlo.kozma@fu-berlin.de}{}{}%TODO mandatory, please use full name; only 1 author per \author macro; first two parameters are mandatory, other parameters can be empty. Please provide at least the name of the affiliation and the country. The full address is optional. Use additional curly braces to indicate the correct name splitting when the last name consists of multiple name parts.

\author{Junqi Tan}{Institut für Informatik, Freie Universität Berlin, Germany}{tanjunqi@zedat.fu-berlin.de}{}{}

\authorrunning{L. Kozma and J. Tan} %TODO mandatory. First: Use abbreviated first/middle names. Second (only in severe cases): Use first author plus 'et al.'

\Copyright{L\'{a}szl\'{o} Kozma and Junqi Tan} %TODO mandatory, please use full first names. LIPIcs license is "CC-BY";  http://creativecommons.org/licenses/by/3.0/

 %TODO mandatory: Please choose ACM 2012 classifications from https://dl.acm.org/ccs/ccs_flat.cfm 

 %TODO mandatory; please add comma-separated list of keywords

\category{} %optional, e.g. invited paper

\relatedversion{} %optional, e.g. full version hosted on arXiv, HAL, or other respository/website
\nolinenumbers %uncomment to disable line numbering

\EventEditors{Anne Benoit, Haim Kaplan, Sebastian Wild, and Grzegorz Herman}
\EventNoEds{4}
\EventLongTitle{33rd Annual European Symposium on Algorithms (ESA 2025)}
\EventShortTitle{ESA 2025}
\EventAcronym{ESA}
\EventYear{2025}
\EventDate{September 15--17, 2025}
\EventLocation{Warsaw, Poland}
\EventLogo{}
\SeriesVolume{351}
\ArticleNo{109}
\begin{document}

\maketitle

\begin{abstract}

For many hard computational problems, simple algorithms that run in time $2^n \cdot n^{O(1)}$ arise, say, from enumerating all subsets of a size-$n$ set. Finding (exponentially) faster algorithms is a natural goal that has driven much of the field of \emph{exact exponential algorithms} (e.g., see Fomin and Kratsch, 2010). 
In this paper we obtain algorithms with running time $O(1.9999977^n)$ on input graphs with $n$ vertices, for the following well-studied problems:
\begin{itemize}
\item \textsc{$d$-Cut}: find a proper cut in which no vertex has more than $d$ neighbors on the other side of the cut; 
\item \textsc{Internal Partition}: find a proper cut in which every vertex has at least as many neighbors on its side of the cut as on the other side; and
\item \textsc{($\upalpha,\upbeta$)-Domination}: given \emph{intervals} $\upalpha,\upbeta \subseteq [0,n]$, find a subset $S$ of the vertices, so that for every vertex $v \in S$ the \emph{number} of neighbors of $v$ in $S$ is from $\upalpha$ and for every vertex $v \notin S$, the number of neighbors of $v$ in $S$ is from $\upbeta$.

\end{itemize}
 Our algorithms are exceedingly simple, combining the \emph{split and list} technique (Horowitz and Sahni, 1974; Williams, 2005) with a tool from computational geometry: \emph{orthogonal range searching} in the moderate dimensional regime (Chan, 2017). 
Our technique is applicable to the decision, optimization and counting versions of these problems and easily extends to various generalizations with more fine-grained, \emph{vertex-specific} constraints, as well as to \emph{directed}, \emph{balanced}, and other variants. 

Algorithms with running times of the form $c^{n}$, for $c<2$, were known for the first problem only for constant $d$, and for the third problem for certain special cases of $\upalpha$ and $\upbeta$; for the second problem we are not aware of such results. 

\end{abstract}

\section{Introduction}
\label{sec:Introduction}

For many hard computational problems, simple algorithms that run in time $2^n \cdot n^{O(1)}$ can be obtained, say, by enumerating all subsets of a size-$n$ set. Obtaining speedups of a factor exponential in $n$ is a natural goal that has driven much of the field of \emph{exact exponential algorithms}, motivating the development of powerful and general algorithmic techniques.   

Landmark successes include \textsc{Independent Set}~\cite{TarjanT77}, \textsc{Subset Sum}~\cite{HorowitzS74}, \textsc{Hamiltonian Cycle}~\cite{Bjorklund14}, \textsc{Max Cut}~\cite{Williams05}, \textsc{$3$-CNF SAT}~\cite{Schoning99}, to name just a few. While all NP-hard, these problems admit algorithms running in time $c^n$ with $c<2$, significantly faster than simple enumeration or dynamic programming over subsets.\footnote{For the problems we mention, $n$ refers to the most natural problem-specific input parameter, e.g., the number of vertices for graph problems, the number of variables for \textsc{CNF SAT}, or the number of elements for \textsc{Set Cover} or \textsc{Subset Sum}.}  We refer to the textbook of Fomin and Kratsch~\cite{FominK10} for a broad overview of such results and techniques, as well as to the earlier, influential survey of Woeginger~\cite{woeginger}.

From a practical point of view, replacing a runtime of  $2^n$ by (say) $1.999^n$ is, of course, rarely consequential, especially if the latter algorithm is more complicated. Yet, such improvements often point at problem-specific algorithmic structure that can be further exploited. Also, the search for algorithms that break such ``triviality barriers'' has often led to the invention of powerful algorithmic techniques that have found broader applications; examples include \emph{branch and reduce}, \emph{measure and conquer}, or \emph{split and list}.

Yet, the set of techniques known to lead to such speedups is not large and several important problems seem to resist improvements beyond $2^n$ (\textsc{Graph Coloring}, \textsc{CNF SAT}, \textsc{Set Cover}, \textsc{TSP} are prominent examples). The hypothesis that no substantially faster algorithm exists for \textsc{CNF SAT} (the \emph{SETH Conjecture}~\cite{Seth}) has become a cornerstone of fine-grained complexity theory, and other barriers are similarly conjectured (e.g., the \emph{Set Cover Conjecture}~\cite{SeCoCo1, SeCoCo2}). There are also many lesser known problems ``stuck'' at $2^n$, where the existing techniques do not seem applicable. Problems where such a barrier was recently bypassed include, e.g., various scheduling~\cite{CyganPPW14, NederlofSW25}, bin packing~\cite{NederlofPSW23},  clustering~\cite{clustering1, clustering2}, multicut~\cite{multicut} and multiway cut~\cite{multiway1, multiway2} problems, to mention just a few.  

\subparagraph{Restricted cut problems.} In this paper, we mainly consider \emph{cut} problems in graphs, with certain natural restrictions on how vertices may interact with the cut. We focus on decision problems (whether a cut of the required form exists); in all cases, an actual cut can be constructed with minimal overhead.

\medskip

Our first problem is \framebox{\textsc{$d$-Cut}}: given an undirected graph $G = (V,E)$ and an integer $d$, find a bipartition $V = V_L \cup V_R$, where $V_L, V_R$ are nonempty disjoint sets, so that each vertex $v \in V$ has at most $d$ neighbors on the other side of the partition. Precisely, denoting by $N(v)$ the (open) set of neighbors of $v$, we require $|N(v) \cap V_R| \leq d$ for all $v \in V_L$ and $|N(v) \cap V_L| \leq d$ for all $v \in V_R$. 
The value $d$ is arbitrary, in particular, it may depend on $n$.  

The problem can be seen as a natural \emph{min-max} variant of the famous \textsc{Min Cut} problem. While \textsc{Min Cut} is polynomial time solvable, \textsc{$d$-Cut} is NP-hard, even for $d=1$~\cite{Chvatal84}. 

The \textsc{$d$-Cut} problem was studied by Gomes and Sau~\cite{GomesSau} as a natural generalization of the well-known \textsc{Matching Cut} problem (the case $d=1$). Graphs admitting a matching cut are known as \emph{decomposable} and have been extensively studied, e.g., see Graham~\cite{graham1970primitive}, Chv\'{a}tal~\cite{Chvatal84}, and Bonsma~\cite{Bonsma}. For \textsc{Matching Cut}, Komusiewicz, Kratsch, and Le~\cite{KKL20} give an $O(1.3071^n)$ time algorithm, using a reduction to \textsc{CNF SAT}. For the general \textsc{$d$-Cut} problem, Gomes and Sau obtain a runtime of the form $O(c_d^n)$ where $c_d$ grows at least as $(2^d-1)^{\frac{1}{d}}$. As such, the base is smaller than $2$ for \emph{fixed} $d$, but cannot, in general, be bounded away from $2$; the existence of an algorithm with runtime $O((2-\varepsilon)^n)$ for $\varepsilon > 0$ was left open by Gomes and Sau. In fact, the \textsc{SAT}-based algorithm of Komusiewicz also applies to \textsc{$d$-Cut}, resulting in a polynomial number of \textsc{$(d+2)$-CNF SAT} instances with $n$ variables. Under the SETH conjecture~\cite{Seth}, the running time of this approach also cannot be bounded as $O((2-\varepsilon)^n)$.   

We note that the related \textsc{Max Cut} problem can be solved in time $O(1.73^n)$ by reduction to fast matrix multiplication~\cite{Williams05}, this technique however, seems not to extend to $d$-\textsc{Cut}. 

\medskip

A second general cut problem we consider is \framebox{\textsc{$(\upalpha,\upbeta)$-Domination}}. Here, we seek a bipartition $V = V_L \cup V_R$, where 
$|N(v) \cap V_L| \in \upalpha$ for all $v \in V_L$ and $|N(v) \cap V_L| \in \upbeta$ for all $v \in V_R$. Note that $\upalpha$ and $\upbeta$ are subsets of the set $\{0,1,\dots,n\}$.

This formulation, introduced by Telle~\cite{Telle}, generalizes many natural structures and problems in graphs, such as \textsc{Independent Set}, \textsc{Dominating Set}, \textsc{Induced Matching}, etc., by setting $\upalpha$ and $\upbeta$ to particular values (we refer to~\cite{Telle, FGK} for a list of special cases). While the fully general case seems hopeless, it is perhaps most natural to require $\upalpha$ and $\upbeta$ to be intervals, i.e., of the form $\upalpha = [\upalpha',\upalpha''], \upbeta = [\upbeta',\upbeta'']$. This already captures most, if not all of the interesting special cases. Notice, however, that the problem does not strictly generalize \textsc{$d$-Cut}; while setting $\upbeta=[0,d]$ enforces the degree constraint on $V_R$, the admissible degrees for vertices in $V_L$ depend on their original degrees in $G$.

The problem was studied from the point of view of exponential algorithmics by Fomin, Golovach, Kratochv\'{i}l, Kratsch, and Liedloff~\cite{FGK}. They obtain algorithms with runtime $c^n$ for $c<2$ in the special case $|\upalpha| = |\upbeta| = 1$, as well as in some cases where $\upalpha$ and $\upbeta$ are not necessarily intervals, e.g., when $|\upalpha| + |\upbeta| = 3$ or when $\upalpha$ and $\upbeta$ are generated by arithmetic sequences; these latter cases fall outside of our definition. 
The same authors, in a different paper~\cite{recharge} also consider other special cases. These require some combination of $\alpha$ and $\beta$ being disjoint, finite, or successor-free (i.e., not containing consecutive integers). As such, the conditions are not comparable with our requirement for $\alpha$ and $\upbeta$ to be intervals; in our case, $\upalpha$ and $\upbeta$ may overlap arbitrarily, may be infinite, but cannot have gaps. 

\medskip

The third problem we consider is \framebox{\textsc{Internal Partition}}. Here, we seek a bipartition $V = V_L \cup V_R$, where $V_L$ and $V_R$ are nonempty and every vertex has at least half of its neighbors on its own side of the partition. Formally, for all $v \in V_L$ we have
$|N(v) \cap V_L| \geq |N(v) \cap V_R|$, and for all $v \in V_R$ we have
$|N(v) \cap V_L| \leq |N(v) \cap V_R|$. 

Such a partition, called an \emph{internal partition}, arises in many settings (in the literature the terms \emph{friendly-}, \emph{cohesive-}, or \emph{satisfactory-} partition are also used). 
Determining whether a graph admits an internal partition is NP-hard~\cite{internal_survey}; this is in contrast to \emph{external partitions} (where every vertex has at least half of its neighbors on the other side), which always exist. Research has mostly focused on conditions that guarantee the existence of an internal partition, e.g., see Thomassen~\cite{thomassen}, Stiebitz~\cite{stiebitz}, and the survey of Bazgan, Tuza, and Vanderpooten~\cite{internal_survey}. A long-standing conjecture is that every sufficiently large $r$-regular graph has an internal partition; e.g., see Ban and Linial~\cite{BanLinial} and Linial and Louis~\cite{linial2020asymptotically} for recent partial results. We are not aware of algorithms with runtime of the form $O((2-\varepsilon)^n)$ for deciding the existence of an internal partition. 

All three problems can trivially be solved in time $2^n \cdot n^{O(1)}$. Our main result improves this for all three problems.

\begin{theorem}
The \textsc{$d$-Cut}, \textsc{$(\upalpha, \upbeta)$-Domination}, and \textsc{Internal Partition} problems can be solved in time $O(1.9999977^n)$.
\end{theorem}

While we state our results for the decision problems (does a bipartition with the given property exist?), we can also easily extend them to the
\emph{optimization} versions of the problems (maximize or minimize the left side $|V_L|$ of the cut while satisfying the constraint), and to the \emph{counting} versions (count the number of cuts with the given property). The optimization view is most natural for \textsc{$(\upalpha,\upbeta)$-Domination}, where special cases include the task of finding e.g., a maximum independent set, a minimum dominating set, etc.~(of course, for some of the well-known special cases faster algorithms already exist).

Additional constraints on the cuts (e.g., requiring $V_L$ to be of a certain size, say $|V_L| = n/2$), or making the edges directed (and extending the conditions accordingly to in- or out-neighborhoods) can easily be accommodated, as will be clear later. 

Our algorithms also extend naturally to more fine-grained, \emph{vertex-specific} conditions. For \textsc{$d$-Cut}, this could mean setting an upper bound $d_v$ for each vertex $v$ on the number of neighbors across the cut, and possibly, a vertex-specific \emph{lower bound} on the same quantity. For \textsc{$(\upalpha,\upbeta)$-Domination}, we could set specific intervals $[\upalpha_v, \upbeta_v]$ for each vertex $v$; these may even depend on the degree of $v$ in the input, or on other properties of the graph. 

\subparagraph{Overview of our approach.} In spite of the unwieldy bound on the running time, our approach is conceptually exceedingly simple, and all three problems are tackled in essentially the same way. 

Our starting point is the \emph{split and list} technique originally used by Horowitz and Sahni~\cite{HorowitzS74} for \textsc{Subset Sum}: Given a set $X = \{x_1, \dots, x_n\}$, find a subset of $X$ whose elements add up to a given value $t$.

The technique works by splitting $X$ into $X_A = \{x_1, \dots, x_{\lfloor n/2 \rfloor}\}$ and $X_B = X \setminus X_A$. Then, the sets $\mathcal{S}_A$, resp., $\mathcal{S}_B$ of the possible sums given by subsets of $X_A$, resp., $X_B$ are computed. The problem now amounts to searching, for all $x \in \mathcal{S}_A$, for a matching pair $t-x \in \mathcal{S}_B$. This can be solved in time $O(n)$ by binary search if $\mathcal{S}_B$ is sorted, which can be achieved by an $O(n2^{n/2})$ time preprocessing. An overall runtime of $O(n2^{n/2})$ results.

We can view the sorting/searching part of the algorithm as storing $\mathcal{S}_B$ in an efficient comparison-based dictionary (say, a balanced binary search tree). Our extension of the technique amounts to replacing this dictionary with a more sophisticated multidimensional search structure. 

In particular, we use the \emph{orthogonal range search} data structure of Chan~\cite{chan2019orthogonal}. Such a data structure stores a set of $N$ vectors from $\mathbb{R}^d$. A \emph{dominance range query} $q \in \mathbb{R}^d$ returns the number (or the list) of vectors stored that are upper bounded by $q$ on every coordinate. Precisely, a stored vector $y = (y_1, \dots, y_d)$ is reported for query vector $q = (q_1, \dots, q_d)$ if $y_i \leq q_i$ for all $i \in \{1,\dots,d\}$.

In most applications of range searching, the dimension $d$ is assumed constant, however, Chan's data structure allows for $d$ that is polylogarithmic in $N$. As we only need $d \in O(\log{N})$, we state the result in this form. 
Note that dominance range queries can be reduced to \emph{orthogonal range queries}, where each query is specified by an axis-parallel hypercube (box), and vice versa (by doubling the number of dimensions). In our application, we always formulate our queries as dominance range queries. 

A small remark on the computational model is in order. While the data structure is defined for vectors in $\mathbb{R}^d$, assuming a real RAM with constant time operations, in our application all vector coordinates are integers in $\{-2n, \dots, 2n\}$, our results thus hold in more realistic models such as the word RAM, or even if we account for the number of bit-operations. 

The result we rely on takes the following form.

\begin{lemma}[Chan~\cite{chan2019orthogonal}]\label{testenv-lemma2}
We can preprocess $N$ vectors from $\mathbb{R}^d$ for $d \leq c \log_2{N}$, and store them in a data structure that serves a sequence of $N$ dominance range queries in a total time (including the preprocessing) $N^{2-\varepsilon_c}$, where $\varepsilon_c>0$ depends only on $c$.
\end{lemma}

We use this geometric data structure to solve the mentioned graph cut problems similarly to the split and list technique sketched above. We split the set of vertices in two equal parts, and form all possible bipartitions of both parts. Combining two such bipartitions into one requires checking whether they are ``compatible'', i.e., whether their union  satisfies the degree constraints. This can, in all three cases, be verified with dominance range queries on appropriately preprocessed degree vectors.  We precisely state and prove our results in Section~\ref{sec:main}.

In Section~\ref{sec:constant} we derive an exact constant $\varepsilon_c$ for Lemma~\ref{testenv-lemma2} in our setting; this is not easy to read off directly from previous work, as results are stated in asymptotic notation, e.g., $\varepsilon_c \in 1/O(\log{c})$ where $c$ is allowed to be (slightly) superconstant. In our application, as we will show, $c=16$ and $\varepsilon_c \geq 3\cdot 10^{-6}$ are feasible, which for $N=2^{n/2}$ yield the stated runtime. 

Chan~\cite{chan2019orthogonal} gives several different implementations that result in a bound of the form given in Lemma~\ref{testenv-lemma2}. We work out the constant in the exponent for the variant that appears the simplest~\cite[\S\,2.1]{chan2019orthogonal}. It is a randomized, ``purely combinatorial'' implementation, i.e., not using fast matrix multiplication or other algebraic tools. The data structure 
allows for an \emph{online} sequence of queries. As in our setting the queries are \emph{offline} (i.e., known upfront), we could use one of the more efficient offline, as well as deterministic versions from~\cite{chan2019orthogonal}, to obtain a similar, possibly better constant; we refrain from this to keep the presentation simple.

While the use of a geometric data structure may seem surprising in a graph theoretic context (its implementation involves an intricate lopsided recursion reminiscent of k-d-trees), Chan's technique did have previous graph-applications, although mostly in the polynomial-time regime, e.g., for the all pairs shortest paths problem~\cite{chan2019orthogonal}. For a different flavor of applying orthogonal range searching (in constant dimensions) to graph algorithms, see~\cite{CabelloK09}. 

After presenting our results (Section~\ref{sec:main}) and deriving 
 a precise constant bound for the base of the running time (Section~\ref{sec:constant}), in Section~\ref{sec:conclusion} we give some final thoughts. 

\section{Results}\label{sec:main}
We use standard graph notation. A bipartition (cut) of an undirected graph $G=(V,E)$ is a partitioning $(V_L,V_R)$ of the set of vertices, i.e., $V = V_L \cup V_R$ and $V_L \cap V_R = \emptyset$. A \emph{proper} bipartition is one where both $V_L$ and $V_R$ are nonempty. 
We denote the open neighborhood of a vertex $v$ by $N(v)$, i.e., $N(v) = \{u\in V : uv \in E\}$. For convenience, for a set $A \subseteq V$ we denote $N_A(v) = N(v) \cap A$. 

\subsection{\textsc{Internal partition}}\label{sec21}
We start with the \textsc{Internal Partition} problem whose solution is the simplest. We first formally define the problem. 

\begin{tcolorbox}
\textsc{\textsc{Internal partition}}\\
\textbf{Input:}  An undirected graph $G = (V , E)$.\\
\textbf{Output:} Is there a proper bipartition $(V_L,V_R)$ of $V$ such that
\begin{itemize}
\item $|N_{V_L}(v)| \geq |N_{V_R}(v)|$ for all $v \in V_L$,
\item $|N_{V_R}(v)| \geq |N_{V_L}(v)|$ for all $v \in V_R$.
\end{itemize}
In words, every vertex has at least half of its neighbors in its own part. 
\end{tcolorbox}

\subparagraph{Algorithm.} We now describe our algorithm for \textsc{Internal Partition}. Start by choosing an arbitrary subset $V_A$ of $V$ of size $\lfloor n/2 \rfloor$ and let $V_B = V \setminus V_A$. For ease of notation, assume $V_A=\{ v_1, \dots, v_{\lfloor n/2 \rfloor}\}$, $V_B=\{ v_{\lfloor n/2 \rfloor +1}, \dots, v_{n}\}$.  

We consider every possible bipartition of $V_A$ and every possible bipartition of $V_B$. Our goal is to identify a ``matching pair'', i.e., a bipartition $(S,R)$ of $V_A$ and a bipartition $(S',R')$ of $V_B$ so that $(S\cup S', R \cup R')$ is a valid bipartition of $V$, satisfying the given degree constraints. 

To achieve this, we encode every proper bipartition $(S,R)$ of $V_A$ as a vector $q = q_S$ and every proper bipartition $(S',R')$ as a vector $p = p_{S'}$, both of dimension $2n$. Our encoding is meant to ensure that $(S \cup S', R \cup R')$ is a feasible solution for \textsc{Internal Partition} if and only if $q \geq p$, i.e., vector $q$ dominates vector $p$. Intuitively, $p$ is a \emph{data vector} stored in the data structure, and $q$ is a \emph{query vector}.

We define $q_S = (q_1^1,\dots,q_{n}^1,q_1^2,\dots,q_n^2)$ as follows. For all $i = 1,\dots,n$:
\begin{itemize}
\item if $v_i \in S$, then $q_i^1=|N_{S}(v_i)|-|N_{R}(v_i)|$, and $q_{i}^2=n$, 
\item if $v_i \in R$, then $q_i^1=n$, and $q_{i}^2=|N_{R}(v_i)|-|N_{S}(v_i)|$, 
\item if $v_i \in V_B$, then $q_i^1=|N_{S}(v_i)|-|N_{R}(v_i)|$, and $q_{i}^2=|N_{R}(v_i)|-|N_{S}(v_i)|$.
\end{itemize}

Intuitively, the entries $q_i^1$ and $q_i^2$ store the ``excess degree'' of $v_i$ to its own part. The two groups of entries correspond to the two inequalities in the definition of the problem. %, resulting from the current cut, when $v_i \in S$, resp., $v_i \in R$. 
If $v_i \in V_A$, then its assignment to either set \(S \subseteq V_L\) or \(R \subseteq V_R\) is already determined, so only one of the two inequalities is relevant to $v_i$. Accordingly, we set one of $q_i^1$ and $q_i^2$ to the correct excess degree, while we set the other to a large value ($n$) that ensures dominance. 
If $v_i \in V_B$, then its assignment to $V_L$ or $V_R$ is not yet determined, and thus, we compute the excess degrees to both parts.

Symmetrically, for every proper bipartition $(S',R')$ of $V_B$, we create a vector $p_{S'} = (p_1^1,\dots,p_{n}^1$, $p_1^2,\dots,p_n^2)$ as follows. For all $i = 1,\dots,n$:
\begin{itemize}
\item if $v_i\in S'$, then $p_i^1=|N_{R'}(v_i)|-|N_{S'}(v_i)|$, and $p_{i}^2=-n$, 
\item if $v_i \in R'$, then $p_i^1=-n$, and $p_{i}^2=|N_{S'}(v_i)|-|N_{R'}(v_i)|$,
\item if $v_i \in V_A$, then $p_i^1=|N_{R'}(v_i)|-|N_{S'}(v_i)|$, and $p_{i}^2=|N_{S'}(v_i)|-|N_{R'}(v_i)|$.
\end{itemize}

The algorithm now stores all vectors $p_{S'}$ in a data structure and queries, for each vector $q_S$, whether it dominates any of the stored vectors. If a dominating pair is found, then the output is yes, otherwise no.

Both the number of stored vectors and the number of queries is at most $N = 2^{\lceil \frac{n}{2} \rceil}$. The construction of the vectors clearly takes time polynomial in $n$ per vector. 

Of course, comparing each pair directly would yield no runtime improvement. However, by applying the data structure of Lemma~\ref{testenv-lemma2}, we can solve the problem in time $O((2^{1-\varepsilon_c/2})^n)$ time, where $c=4$, since the dimension is $2n = 4 \log_2{N}$.

We required $(S,R)$ and $(S',R')$ to be \emph{proper} bipartitions of $V_A$ and $V_B$, to avoid a possible trivial solution where one part is empty. This means that we need to check the cases $S=\emptyset$ or $R=\emptyset$ against all proper bipartitions of $V_B$ and the cases  $S'=\emptyset$ or $R'=\emptyset$ against all proper bipartitions of $V_A$. These special cases require time $2^{n/2} \cdot n^{O(1)}$, which is absorbed in the overall runtime.  

\subparagraph{Correctness.} We must show that $V$ admits a feasible bipartition if and only if there is some $q_S$ that dominates some $p_{S'}$. 

Suppose that $(V_L, V_R)$ is a feasible bipartition, and let $V_A,V_B$ be the algorithm's choice, so $S = V_A \cap V_L$ and $S' = V_B \cap V_L$, and $R = V_A \cap V_R$ and $R' = V_B \cap V_R$ are among the sets considered. The case when one of $S,S',R,R'$ is empty is handled separately, and a solution can clearly be found. Otherwise, let $q_i^k$, $p_i^k$, for $i \in \{1, \dots, n\}$ and $k \in \{1,2\}$ denote the entries of $q_S$ and $p_{S'}$, as defined above.

If $v_i \in S \subseteq V_L$, or if $v_i \in S' \subseteq V_L$, then $q_i^1 - p_i^1 = |N_S(v_i)| - |N_R(v_i)| + |N_{S'}(v_i)| - |N_{R'}(v_i)| = |N_{V_L}(v_i)| - |N_{V_R}(v_i)|$. This must be nonnegative from the feasibility condition. For the second part, if $v_i \in S \subseteq V_L$, then $q_{i}^2 = n$, and $p_{i}^2 = |N_{S'}(v_i)| - |N_{R'}(v_i)|$, which clearly yields $q_{i}^2 \geq p_{i}^2$. If $v_i \in S' \subseteq V_L$, then $q_{i}^2 - p_{i}^2 = |N_{R}(v_i)| - |N_{S}(v_i)| + n \geq 0$.

If $v_i \in R \subseteq V_R$, or if $v_i \in R' \subseteq V_R$,  then $q_{i}^2 - p_{i}^2 = |N_R(v_i)| - |N_S(v_i)| - |N_{S'}(v_i)| + |N_{R'}(v_i)| = |N_{V_R}(v_i)| - |N_{V_L}(v_i)|$, again, nonnegative from the feasibility condition. For the first part, if $v_i \in R \subseteq V_R$, then $q_{i}^1 = n$, and $p_{i}^1 = |N_{R'}(v_i)| - |N_{S'}(v_i)|$, which yields $q_{i}^1 \geq p_{i}^1$. If $v_i \in R' \subseteq V_R$, then $q_{i}^1 - p_{i}^1 = |N_{S}(v_i)| - |N_{R}(v_i)| + n \geq 0$.

\medskip

Conversely, if $q_S \geq p_{S'}$, then from the corresponding $q_i^1 \geq p_i^1$ we obtain $N_{S \cup S'}(v_i) \geq N_{V \setminus (S \cup S')}(v_i)$ for all $v_i \in S \cup S'$. From $q_i^2 \geq p_i^2$ we obtain $N_{V \setminus (S \cup S')}(v_{i}) \geq N_{S \cup S'}(v_i)$ for all $v_i \in V \setminus (S \cup S')$. This means that $(V_L, V_R) = (S \cup S', V \setminus (S \cup S'))$ is a feasible solution.

\if 0
We assume that the set \( V \) is partitioned into \( V_L \) and \( V_R \), where $V_L=S_0\cup S_1$ and \(V_R=V \setminus (S_{0} \cup S_{1})\).
The vector \( (p_1, p_2, \ldots, p_{2n}) \) encodes potential assignments of \(v_j\), where index \(i \leq n\) indicates assignment to \(V_L\), and \(i > n\) indicates assignment to \(V_R\).
Furthermore, if \(i \in [1,n]\), then \(p_i\) represents the difference between the number of neighbors that $v_i$ has in $S_0$ and in $V_0\setminus S_0$; Otherwise, \(p_i\) represents the difference between the number of neighbors that $v_i$ has in $V_0\setminus S_0$ and in $S_0$. If a vertex \(v\) belongs to the subset \(V_0\), its assignment to either set \(V_L\) or \(V_R\) is predetermined. Therefore, we only need to compute the neighbor counts for $p_i$ or $p_{n+i}$, while the other is filled with an auxiliary value. Therefore, there is a  feasible solution for \textsc{Internal partition} iff there exist $(p_1,p_2,...,p_{2n})$ and $(q_1,q_2,...,q_{2n})$ such that $p_i+q_i\ge 0$ for all $i\in [1,2n]$. The theorem is proved.
\textit{Running time} \quad The number of vectors is $2\times 2^{n/2}$. For every vector, we can construct it in poly($n$) time. By applying Lemma~\ref{testenv-lemma2}, we can answer \textsc{Internal partition} in $O^*((2^{1-\varepsilon_c /2})^n)$ time, where $c=4$.
\fi

\subsection{\textsc{$d$-Cut} and \textsc{$(\upalpha, \upbeta)$-Domination}}

Let us formally state our other two problems.

\begin{tcolorbox}
\textsc{\textsc{$d$-Cut}}\\
\textbf{Input:}  An undirected graph $G = (V , E)$ and a nonnegative integer $d$.\\
\textbf{Output:} Is there a proper bipartition $(V_L,V_R)$ of $V$ such that 
\begin{itemize}
\item $|N_{V_L}(v)| \leq d$ for all $v \in V_R$,
\item $|N_{V_R}(v)| \leq d$ for all $v \in V_L$.
\end{itemize}
In words, every vertex is incident to at most $d$ edges that cross $(V_L, V_R)$.
\end{tcolorbox}

\begin{tcolorbox}
\textsc{\textsc{$(\upalpha,\upbeta)$-Domination}}\\
\textbf{Input:}  An undirected graph $G = (V , E)$ and intervals $\alpha, \beta \subseteq [0,n]$.\\
\textbf{Output:} Is there a proper bipartition $(V_L,V_R)$ of $V$ such that 
\begin{itemize}
\item $|N_{V_L}(v)| \in \alpha$ for all $v\in V_L$,
\item $|N_{V_L}(v)| \in \beta$ for all $v\in V_R$.
\end{itemize}
\end{tcolorbox}

We next introduce a general problem formulation that allows for vertex-specific constraints and that extends both problems. 

\begin{tcolorbox}
\textsc{\textsc{Interval-Constrained Cut}}\\
\textbf{Input:}  An undirected graph $G = (V , E)$ and nonnegative integers $a'_v,a''_v,b'_v,b''_v$, $c'_v,c''_v$, $d'_v,d''_v$ for all $v$.\\
\textbf{Output:} Is there a proper bipartition $(V_L,V_R)$ of $V$ such that
\begin{itemize}
\item $|N_{V_L}(v)| \in [a'_v,a''_v]$ for all $v \in V_L$,
\item $|N_{V_R}(v)| \in [b'_v,b''_v]$ for all $v \in V_L$,
\item $|N_{V_R}(v)| \in [c'_v,c''_v]$ for all $v \in V_R$,
\item $|N_{V_L}(v)| \in [d'_v,d''_v]$ for all $v \in V_R$.

\end{itemize}
\end{tcolorbox}

The problem generalizes \textsc{$d$-Cut}, by setting $[a'_v,a''_v] = [c'_v,c''_v] = [0,n]$, i.e., ``don't care'', and setting $[b'_v,b''_v] = [d'_v,d''_v] = [0,d]$, i.e., the required degree constraint. 

The problem also generalizes \textsc{$(\upalpha, \upbeta)$-Domination}, for intervals $\upalpha = [\upalpha',\upalpha'']$ and $\upbeta= [\upbeta',\upbeta'']$, by setting $[b'_v,b''_v] = [c'_v,c''_v] = [0,n]$, i.e., ``don't care'', and setting $[a'_v,a''_v] = [\upalpha',\upalpha'']$ and $[d'_v,d''_v] = [\upbeta',\upbeta'']$, i.e., the required degree constraint.

\medskip

\subparagraph{Algorithm.} We now describe our solution for \textsc{Interval-Constrained Cut}, which is very similar to the earlier one (only slightly more complicated, since the problem definition involves $8$ inequalities instead of $2$).

Again, we choose an arbitrary subset $V_A$ of $V$ of size $\lfloor n/2 \rfloor$ and let $V_B = V \setminus V_A$, denoting $V_A=\{ v_1, \dots, v_{\lfloor n/2 \rfloor}\}$, $V_B=\{ v_{\lfloor n/2 \rfloor +1}, \dots, v_{n}\}$.

We encode every proper bipartition $(S,R)$ of $V_A$ as a vector $q_S$ and every proper bipartition $(S',R')$ as a vector $p_{S'}$, where both vectors have dimension $8n$. (Bipartitions where one part is empty are handled separately, in the same way as in Section~\ref{sec21}.) Additionally, we encode the problem constraints as a vector $r$, also of dimension $8n$. The encoding ensures that $q_S$ dominates $p_{S'}+r$ if and only if $(S\cup S', V \setminus (S\cup S'))$ is a feasible solution. 

The entries of vectors $q = q_S$, $p = p_{S'}$, $r$ are indexed as $q_i^k$, $p_i^k$, $r_i^k$ for $i \in \{1, \dots, n\}$ and $k \in \{1,\dots,8\}$, ordered lexicographically according to the index $(k,i)$. 

Intuitively, the entries form $8$ groups, corresponding to the $8$ inequalities of the interval constraints in the problem statement, and $q_i^k \geq p_i^k + r_i^k$ verifies whether the $k$-th inequality holds for vertex $v_i$.

The encoding is as follows. We start with $r$, which is the simplest to state. The entries $r_i^k$ are, for all $i = 1,\dots,n$:

\begin{center}
\begin{tabular}{ |c|c|c|c|c|c|c|c| } 
 $r_i^1$ & $r_i^2$ & $r_i^3$ & $r_i^4$ & $r_i^5$ & $r_i^6$ & $r_i^7$ & $r_i^8$ \\ 
\hline
 $a'_i$ & $-a''_i$ & $b'_i$ & $-b''_i$ & $c'_i$ & $-c''_i$ & $d'_i$ & $-d''_i$\\ 
 \hline
\end{tabular}
\end{center}

We next describe the entries $q_i^k$ of $q_S$ and the entries $p_i^k$ of $p_{S'}$. Recall that $(S,R)$ is a bipartition of $V_A$ and $(S',R')$ is a bipartition of $V_B$. For all $i = 1,\dots,n$:

\begin{table}[h!]
\scalebox{0.935}{
\begin{tabular}{ r|c|c|c|c|c|c|c|c| } 
 $v_i \in $& $q_i^1$ & $q_i^2$ & $q_i^3$ & $q_i^4$ & $q_i^5$ & $q_i^6$ & $q_i^7$ & $q_i^8$ \\ 
\hline
 $S$ & $|N_S(v_i)|$ & $-|N_S(v_i)|$ & $|N_R(v_i)|$ & $-|N_R(v_i)|$ & $2n$ & $2n$ & $2n$ & $2n$\\ 
 $R$ & $2n$ & $2n$ & $2n$ & $2n$ & $|N_R(v_i)|$ & $-|N_R(v_i)|$ & $|N_S(v_i)|$ & $-|N_S(v_i)|$\\
 $V_B$ & $|N_S(v_i)|$ & $-|N_S(v_i)|$ & $|N_{R}(v_i)|$ & $-|N_{R}(v_i)|$ & $|N_R(v_i)|$ & $-|N_R(v_i)|$ & $|N_S(v_i)|$ & $-|N_S(v_i)|$\\
 \hline
\end{tabular}
}
\end{table}

The three rows of the table correspond to the cases where $v_i$ is in $S$, $R$, or $V_B$, showing the value to which $q_i^k$ is set. The $8$ columns distinguish the entries needed for verifying the $8$ inequalities (i.e., the index $k$). Notice that if $v_i \in S$ or $v_i \in R$, then the placement of $v_i$ in $V_L$ or $V_R$ is determined and only $4$ of the $8$ inequalities in the definition are applicable to $v_i$. For these cases we set $q_i^k$ to the relevant degree of $v_i$ (with plus sign where we have a lower bound, and minus sign where we have an upper bound). For the inequalities that do not apply to $v_i$, we set $q_i^k$ to a sufficiently large value of $2n$ which ensures that on the given $(k,i)$ coordinate $q$ dominates $p+r$, as vertex $v_i$ trivially satisfies the condition $k$.

On the other hand, if $v_i \in V_B$, then the placement of $v_i$ within the cut $(V_L,V_R)$ is not yet determined, in this case we account for all degrees, preparing for both cases. 

\medskip

The encoding of $p = p_{S'}$ is symmetric. For all $i = 1,\dots,n$:

%\nolinenumbers
%\scalebox{0.89}{
%\centering
\begin{table}[h!]
\scalebox{0.89}{
\begin{tabular}{ r|c|c|c|c|c|c|c|c| } 
 $v_i \in $& $p_i^1$ & $p_i^2$ & $p_i^3$ & $p_i^4$ & $p_i^5$ & $p_i^6$ & $p_i^7$ & $p_i^8$ \\ 
\hline
 $S'$ & $-|N_{S'}(v_i)|$ & $|N_{S'}(v_i)|$ & $-|N_{R'}(v_i)|$ & $|N_{R'}(v_i)|$ & $-2n$ & $-2n$ & $-2n$ & $-2n$\\ 
 $ R'$ & $-2n$ & $-2n$ & $-2n$ & $-2n$ & $-|N_{R'}(v_i)|$ & $|N_{R'}(v_i)|$ & $-|N_{S'}(v_i)|$ & $|N_{S'}(v_i)|$\\
 $V_A$ & $-|N_{S'}(v_i)|$ & $|N_{S'}(v_i)|$ & $-|N_{R'}(v_i)|$ & $|N_{R'}(v_i)|$ & $-|N_{R'}(v_i)|$ & $|N_{R'}(v_i)|$ & $-|N_{S'}(v_i)|$ & $|N_{S'}(v_i)|$\\
 \hline
\end{tabular}
}
\end{table}
%}
%\linenumbers

We then store all vectors $p_{S'}+r$ in the data structure and execute orthogonal dominance queries for all $q_S$ vectors. A running time of $O((2^{1-\varepsilon_c /2})^n)$ follows similarly to the previous section, with $c=16$, since the dimension is $8n = 16 \log_2{N}$.

\subparagraph{Correctness.} 

We need to verify that there is a feasible solution to the \textsc{Interval-Constrained Cut} problem if and only if there is a $q_{S}$ vector that dominates some $p_{S'} + r$ vector. 

Suppose first that a feasible solution $(V_L,V_R)$ exists and let $S = V_A \cap V_L$ and $S' = V_B \cap V_L$. We show that $q_S \geq p_{S'} + r$, so the algorithm correctly finds the solution. 

We verify this dominance relation on the entries $q_i^1$, $p_i^1$, $r_i^1$; the calculations for the cases $k=2,\dots,8$ are entirely analogous and thus omitted. 

If $v_i \in R$ or $v_i \in R'$ then either $q_i^1 = 2n$ or $p_i^1 = -2n$, and in both cases $q_i^1 \geq p_i^1 + r_i^1$, since $r_i^1 = a_i' \leq n$. Thus, the inequality holds.

If $v_i \in S$ or if $v_i \in S'$, then $q_i^1 = |N_S(v_i)|$ and $p_i^1 = -|N_{S'}(v_i)|$. Thus $q_i^1 \geq p_i^1 + r_i^1$ is equivalent to $|N_{S}(v_i)| + |N_{S'}(v_i)| \geq a_i'$. Since the left side equals $|N_{V_L}(v_i)|$, the inequality follows from the feasibility of $(V_L,V_R)$.

\medskip

Conversely, suppose that the algorithm identifies vectors $q_S$ and $p_{S'}$ for which $q_S \geq p_{S'} + r$ holds. In particular, if $v_i \in S \cup S'$, then $q_i^1 \geq p_i^1 + r_i^1$, which amounts to $|N_S(v_i)| + |N_{S'}(v_i)| \geq a_i'$, yielding $|N_{S \cup S'}(v_i)| \geq a_i'$. This means that a partition $(V_L, V_R)$ with $V_L = S \cup S'$ satisfies the first inequality of the problem statement. If $v_i \notin S \cup S'$, then the first inequality does not apply to $v_i$, and thus trivially holds. The remaining inequalities follow analogously from the cases $k=2,\dots,8$. Thus, a feasible solution $(S \cup S', V \setminus (S \cup S'))$ exists.

\subparagraph{Remarks.}

If only the \textsc{$d$-Cut} problem is being solved, then one can encode the partitions with only $2$ entries per vertex instead of $8$, similarly to \textsc{Internal Partition}, yielding a (small) saving in runtime. 
An additional optimization that we omitted is to notice when some partial cuts $(S,R)$ and $(S',R')$ already violate the degree upper bounds; in such cases the corresponding vectors need not be stored, resp., queried.

\subsection{Extensions}
So far we have described how to solve the decision problem, i.e., deciding whether a cut with the given constraints exists. Constructing an actual feasible cut is easy, as the data structure can return for a query vector $q_S$ a single data vector $p_{S'}$ that has led to the positive answer without any overhead. With minor bookkeeping, then $S$ and $S'$ are identified, and the bipartition $(S \cup S', V \setminus (S \cup S'))$ is constructed.

Chan's data structure returns the \emph{number} of data vectors dominated by the query vector. Adding up all these counts yields the total number of solutions without overcounting (notice that every solution is uniquely identified by its intersection with the partition $(V_A,V_B)$ chosen by the algorithm). Note that we must also add the solutions found in the special cases when one of $S,S',R,R'$ is empty.

Finally, if we only look for solutions of a given size, say $|V_L| = t$, then we can encode the size of partial cuts ($|S|$ and $|S'|$) and use two additional dimensions to enforce the inequalities $|S|+|S'| \leq t$ and $|S|+|S'| \geq t$. This does not affect the stated runtimes. If we wish to minimize or maximize the size $|V_L| = t$ for which a solution exists, then we can repeat the procedure for $t = 1,\dots,n-1$ and choose the optimum, with a factor $n$ overhead, which is negligible in our regime.

\section{Obtaining a base constant}
\label{sec:constant}
In this section, for concreteness, we derive a constant value for the base of the running time, without attempting to significantly optimize it.

This results from the analysis of the dominance range searching data structure stated in Lemma~\ref{testenv-lemma2}. We refer to the \emph{online} implementation by Chan described and analyzed in~\cite[\S\,2.1]{chan2019orthogonal}. While the online aspect of the data structure is not strictly needed in our application, the implementation is relatively simple (using a recursive k-d-tree-like approach) and in particular it does not rely on algebraic techniques. While Chan also describes deterministic, offline designs with stronger asymptotic guarantees, they rely on more complicated techniques, and exact constants are more difficult to derive for them (although their running times are also clearly of the form given in Lemma~\ref{testenv-lemma2}). 

Chan upper bounds the cost of a query as $e^{\gamma d}\cdot N^{1-\varepsilon} \cdot (bd \log{N})^{O(1)}$. The parameter $b$ is set as $b = (c/\delta) \log{(c/\delta)}$, where $\delta \in (0,1)$ is a user-chosen parameter, and $\log$ is the base $2$ logarithm.

Recall that in our application $N = 2^{n/2}$ and $d = c\log{N}$, where $c=16$.
Thus, the last factor $(bd \log{N})^{O(1)}$ can be upper bounded by a polynomial in $n$, and thus omitted (polynomial factors are absorbed, if we slightly increase the base of the exponential).

As for the remaining parameters, Chan sets in the analysis $\gamma = 8b \upalpha$, where $\upalpha=\frac{1}{b^4}$, and finally, $\varepsilon = \frac{1}{6b\log{\frac{1}{\upalpha}}} = \frac{1}{24b\log{b}}$.

Thus, (after ignoring a polynomial factor, as mentioned), the cost of $N$ queries is at most:

$$ N^{2-\varepsilon} \cdot e^{\frac{8}{b^3} c\log{N}} = N^{2 - \frac{1}{24b\log{b}}} \cdot e^{\frac{4cn}{b^3}}.$$

We now plug in $c=16$ and $N=2^{n/2}$ and choose $\delta = 0.1$, from which $1171 < b < 1172$ follows.

For the first factor we get an upper bound of $1.999997583^n$ and for the second factor an upper bound of $1.00000003986^n$, making their product less than $1.9999977^n$.

\medskip

It remains to bound the preprocessing cost, i.e., the time needed to build the data structure that stores $N$ vectors. This is given by Chan~\cite[\S\,2.1]{chan2019orthogonal} as $N^{1+O(\delta)}$ multiplied by a factor which resolves to the form $n^{O(\log{n})}$. While this is a \emph{quasi-polynomial} factor, it can still be absorbed by the exponential by a slight rounding up of the exponential base, we thus omit this factor as well. 

The crucial part is the constant hidden in the term $O(\delta)$ in the exponent.
One can observe in the analysis~\cite[\S\,2.1]{chan2019orthogonal} that the term $N^{1+O(\delta)}$ arises as $N \cdot b^{O({d}/{b})}$. The $O()$-notation is used to suppress constant factors in two steps of the analysis. We will make these explicit, stating the bound as $N \cdot b^{\upalpha_1 \cdot \upalpha_2 \cdot ({d}/{b})}$ for constants $\alpha_1, \alpha_2 > 1$ (our notation). 

For our earlier choices of $N$, $b$, $d$, the quantity is upper bounded as $2^{n/2} \cdot 1.0495^{n \cdot \upalpha_1 \cdot \upalpha_2}$.
It remains to resolve the constants $\upalpha_1, \upalpha_2$. 

One of them, $\upalpha_1$, captures a factor of $\sum_{i}{1/b^i}$, absorbed in the $O()$-notation in one step of the analysis. For our choice of $b$, this geometric sum can be upper bounded as $\upalpha_1 \leq \frac{b}{b-1} < 1.00085$.
The other constant, $\upalpha_2$, arises from upper bounding ${d \choose {\lfloor {d}/{b} \rfloor}}$ as $b^{O(d/b)}$. We replace this estimate by the explicit upper bound $b^{\upalpha_2 \cdot (d/b)}$.
This follows from the well-known inequality ${n  \choose k} \leq (\frac{e n}{k})^k$, and since $b>e$,  the constant $\upalpha_2 = 2$ suffices for the inequality to hold.

Putting things together, we obtain a runtime of at most $2^{n/2} \cdot 1.0495^{n \cdot \upalpha_1 \cdot \upalpha_2} < 1.558^n$, which is within our stated bounds.

\section{Conclusion}\label{sec:conclusion}
In this paper we gave an algorithm that solves a family of constrained cut problems, combining the split and list technique with a sophisticated geometric data structure that allows orthogonal range queries on degree vectors. 
Our method can also solve, with minor adaptation, the following vector-based generalization of \textsc{Subset Sum} in similar $(2-\varepsilon)^n$ running time: given a set of $n$ vectors of dimension linear in $n$, find a subset of vectors whose sum is within a target box $t$.

Our overall approach is conceptually very simple. One may complain that this simplicity is illusory, since the complexity is all hidden inside the data structure; to which we would respond: \emph{precisely, that is the point of the method}. In all fairness, one could attempt to ``unroll'' the data structure implementation to see the algorithm in more elementary steps.
We instead propose a different question: can we replace the orthogonal range search structure by different kinds of data structures, to extend \emph{split and list} to further algorithmic problems?

The space usage of Chan's data structure is, with careful optimization, near-linear, i.e., of the form $N^{1+\delta}$ where $N = 2^{n/2}$ and $\delta>0$. 
In case of \textsc{Subset Sum}, an original space usage of $2^{n/2}$ has subsequently been reduced~\cite{SS} to $2^{n/4}$ (with a recent small further improvement~\cite{NW}). Can similar space improvements, or a finer time-space-tradeoff (again, similarly to \textsc{Subset Sum}) be obtained for the problems considered in this paper?

%%
%% Bibliography
%%

%% Please use bibtex, 
\newpage

\bibliography{lipics-v2021-sample-article}

\end{document}